\documentclass[conference]{IEEEtran}
\usepackage{color}
\usepackage[english]{babel}
\usepackage{graphicx}
\usepackage{epstopdf}
\usepackage{times}
\usepackage{multirow}
\usepackage[algo2e,linesnumbered]{algorithm2e} 
\usepackage{algorithmic}
\usepackage{algorithm}
\usepackage{psfrag}
\usepackage[shell]{gnuplottex}
\usepackage{mathenv}
\usepackage{array}
\usepackage[T1]{fontenc}
\usepackage{array}
\usepackage{mdwmath}
\usepackage{multicol}
\usepackage{amsmath}
\usepackage{amssymb}
\usepackage{float}
\usepackage{url}
\newcommand{\subparagraph}{}
\usepackage{lipsum}
\usepackage{titlesec}
\titlespacing\section{0pt}{6pt plus 2pt minus 2pt}{6pt plus 2pt minus 2pt}
\titlespacing\subsection{0pt}{3pt plus 2pt minus 2pt}{3pt plus 2pt minus 2pt}

\usepackage{mathtools}

\usepackage{dirtytalk}
\usepackage[utf8]{inputenc}
\usepackage{amsthm}

\makeatletter
\newcommand*{\rom}[1]{\expandafter\@slowromancap\romannumeral #1@}
\makeatother

\showoutput
\begin{document}
\title{Inter-WBANs Interference Mitigation Using Orthogonal Walsh Hadamard Codes}
\author{\IEEEauthorblockN{Mohamad Ali\IEEEauthorrefmark{1}, Hassine Moungla\IEEEauthorrefmark{1}, Mohamed Younis\IEEEauthorrefmark{2}, Ahmed Mehaoua\IEEEauthorrefmark{1}}
\IEEEauthorblockA{\IEEEauthorrefmark{1}LIPADE, University of Paris Descartes, Sorbonne Paris Cit\'{e}, Paris, France \\
}{\IEEEauthorrefmark{2}Department of Computer Science and Electrical Engineering, University of Maryland, Baltimore County, United States} \\Email: \{mohamad.ali; hassine.moungla; ahmed.mehaoua\}@parisdescartes.fr; younis@umbc.edu}
\maketitle
\begin{abstract}
A Wireless Body Area Network (\textit{WBAN}) provides health care services. The performance and utility of \textit{WBAN}s can be degraded due to interference. In this paper, our contribution for co-channel interference mitigation among coexisting \textit{WBAN}s is threefold. First, we propose a distributed orthogonal code allocation scheme, namely, \textit{OCAIM}, where, each \textit{WBAN} generates \textit{sensor interference lists (SILs)}, and then all sensors belonging to these lists are allocated orthogonal codes. Secondly, we propose a distributed time reference correlation scheme, namely, \textit{DTRC}, that is used as a building block of \textit{OCAIM}. \textit{DTRC} enables each \textit{WBAN} to generate a virtual time-based pattern to relate the different superframes. Accordingly, \textit{DTRC} provides each \textit{WBAN} with the knowledge about, 1) which superframes and, 2) which time-slots of those superframes interfere with the time-slots within its superframe. Thirdly, we further analyze the success and collision probabilities of frames transmissions when the number of coexisting \textit{WBAN}s grows. The simulation results demonstrate that \textit{OCAIM} outperforms other competing schemes in terms of interference mitigation and power savings.  
\end{abstract}
\section{Introduction}
A \textit{WBAN} is a wireless emerging technology consisting of a coordinator and multiple low power, wearable or implanted tiny sensors for collecting health related data about the physiological state of human body. \textit{WBAN}s are used in many applications such as medical treatment and diagnosis, consumer electronics, sports and military \cite{key15}. For example, these sensors may observe the heart and the brain electrical activities as well as vital signs and parameters like insulin percentage in blood, blood pressure, temperature, etc.

Recently, the \textit{IEEE 802.15.4} standard has proposed new protocols for \textit{WBAN}s and specified an upper limit for the number of \textit{WBAN}s (60 sensors in $6m^3$ and 256 sensors in $3m^3$ space) that properly coexist \cite{key26}. The standard proposes three mechanisms for co-channel interference mitigation, namely, beacon shifting, channel hopping and active superframe interleaving. Nevertheless, there is a great possibility of co-channel interference among coexisting \textit{WBAN}s, e.g., inside hospital's corridor crowded with patients. Hence, communication links may suffer interference, and consequently the performance and quality of service requirements of each individual \textit{WBAN} may quickly degrade. Therefore, interference mitigation is quite necessary for reliable communication in \textit{WBAN}s.

In addition, the highly mobile and resource constrained nature of \textit{WBAN}s make co-channel interference mitigation quite challenging. Such nature makes the allocation of global coordinator to manage multiple \textit{WBAN}s coexistence as well as the application of advanced antenna and power control techniques used in cellular networks unsuitable for \textit{WBAN}s. Moreover, due to the absence of coordination and synchronization among \textit{WBAN}s, the superframes of different nearby \textit{WBAN}s may overlap, and hence, their concurrent transmissions may interfere. More specifically, when two or more sensors of different \textit{WBAN}s access the shared channel at the same time and therefore, the co-channel interference may arise. In this paper, we tackle these issues and contribute the following:
\begin{itemize}
    \item \textit{DTRC}, a scheme for determining which superframes and their corresponding times slots overlap with each others
    \item \textit{OCAIM}, a scheme that allocates orthogonal codes to interfering sensors belonging to sensor interference lists
    \item An analysis of the success and collision probability model for frames transmissions
\end{itemize}
Simulation results and theoretical analysis show that our proposed approach can significantly increase the minimum signal to interference plus noise ratio (SINR) and the energy savings of each \textit{WBAN}. Additionally, our proposed scheme significantly diminishes the inter-\textit{WBAN} interference level and adds no complexity to the sensors as the coordinators are only required to compute and negotiate for orthogonal code assignment. 

The rest of the paper is organized as follows. Section \rom{2} discusses the related work. Section \rom{3} states the system model and covers some preliminaries. Section \rom{4} explains our distributed time reference correlation scheme. Section \rom{5} describes our proposed interference mitigation scheme using orthogonal codes. Section \rom{6} mathematically analyzes the performance of \textit{OCAIM}. Section \rom{7} presents the simulation results. Finally, the paper is concluded in Section \rom{8}.  

\section{Related Work}
Prior work has addressed the problem of co-channel interference in \textit{WBAN}s through spectrum allocation, cooperation, power control, game theory and multiple access schemes. Examples of spectrum allocation techniques include \cite{key20,key16,key25,key60}. However, in \cite{key60}, a prediction algorithm for dynamic channel allocation is proposed where, variations of channel assignment due to \textit{WBAN}s mobility are investigated. In \cite{key16}, a dynamic resource allocation is proposed where, orthogonal channels are allocated for interfering sensors belonging to each pair of \textit{WBAN}s. Movassaghi et al., \cite{key25} have proposed an adaptive interference avoidance scheme that considers sensor-level interference only. Further, the proposed scheme allocates synchronous and parallel transmission intervals and significantly reduces the number of assigned channels. Whereas, our approach considers both sensor- and time-slot-level interference and significantly diminishes the latter as well as better improves the power savings of \textit{WBAN}s. Whilst, Liang et al., \cite{key20} have analyzed the inter-\textit{WBAN} interference using various performance metrics and then, proposed interference detection and mitigation scheme using dynamic channel hopping. 

Meanwhile, other approaches have adopted cooperative communication, game theory and power control to mitigate co-channel interference. Dong et al., \cite{key9} have pursued joint cooperative and power control communication for \textit{WBAN}s coexistence problem. Similarly, in \cite{key62}, intra-\textit{WBAN} interference is mitigated using cooperative orthogonal channels and a contention window extension mechanism. Whereas, the approach of \cite{key30} employs a Bayesian game based power control to mitigate inter-\textit{WBAN} interference by modeling \textit{WBAN}s as players and active links as types of players in the Bayesian model. Other approaches pursued multiple access schemes for interference mitigation. In \cite{key6}, multiple \textit{WBAN}s collaborate to agree on common \textit{TDMA} schedule that avoids the co-channel interference. Whilst, Kim et al., \cite{key61} have proposed distributed \textit{TDMA}-based beacon interval shifting scheme for interference mitigation where, the wakeup period of each \textit{WBAN} coinciding with other \textit{WBAN}s is avoided by employing carrier sense before a beacon transmission. Also, Chen et al., \cite{key3} adopts \textit{TDMA} for scheduling transmissions within a \textit{WBAN} and carrier sensing mechanism to deal with inter-\textit{WBAN} interference.

In this paper, we take a step forward and consider the sensor- and time-slot- levels interference mitigation amongst coexisting \textit{WBAN}s. More specifically, we allocate orthogonal code to each interfering sensor in its assigned time-slot to avoid interference with other \textit{WBAN}s' sensors. Meanwhile, we depend on distributed time reference correlation and time provisioning to determine which superframes overlap with each other. 
\section{System Model and Preliminaries}
\subsection{Model Description and Assumptions}
Let us consider a network composed of \textit{N} coexisting \textit{TDMA}-based \textit{WBAN}s, each consists of up to \textit{K} sensors that transmit their data to a single coordinator denoted by C. All sensors transmit at maximum data rate of \textit{250Kb/s} within the \textit{2.4 GHz} unlicensed international band using the same transmission power (-10 dBm), modulation scheme and average transmitted energy per symbol. Furthermore, we assume that the superframes of different \textit{WBAN}s are neither aligned nor synchronized and may overlap with each other.

\subsection{Interference Lists (\textit{I})}
Let us assume when $k^{th}$ sensor $S_{i,k}$ of \textit{$WBAN_{i}$} is transmitting to its $C_{i}$, in the same time, all other coordinators compute the power received from $S_{i,k}$'s transmitted signal. Let $\delta_{i, j, k}$ denotes the power received from the $k^{th}$ sensor of \textit{$WBAN_{j}$} at the coordinator of \textit{$WBAN_{i}$}. After all transmissions are over in the first round, each $C_{i}$ creates a table consisting of power received from all sensors in the network. Furthermore, we denote the minimum power received within a \textit{$WBAN_{i}$} by $\rho_{i}^{min} = min\{\delta_{i, k = 1, \dots, K}\}$. Therefore, we denote the interference list of \textit{$WBAN_{i}$} by $I_{i}$ and defined as in \textbf{\textbf{eq.}} (\ref{ii}) below.

\small
\begin{equation}\label{ii}
 I_{i} =  \{S_{l, m} | \delta_{i, l, m} > \rho_{i}^{min} - \theta, \forall i \neq l \}
\end{equation}
\normalsize
Where $\theta$ is the interference threshold, afterwards, each $C_{i}$ broadcasts $I_{i}$ to all network coordinators. 

\subsection{Interference Sets (\textit{IS})}
Based on power tables update (using the broadcast among \textit{WBAN}s), each $C_{i}$ verifies which of its sensors impose interference on sensors of other \textit{WBAN}s and which sensors of other \textit{WBAN}s impose interference on its \textit{WBAN}'s sensors. It then creates an interference set denoted by $IS_{i}$ and defined as in \textbf{eq.} (\ref{isi}) below.

\small
\begin{equation}\label{isi}
 IS_{i} = I_{i} \cup \{S_{i,k} | S_{i,k} \in I_{l}, \forall l \neq i\}
\end{equation}
\normalsize
\subsection{Cyclic Orthogonal Walsh Hadamard Codes \textit{(COWHC)}}
In this section, we provide a brief overview of cyclic orthogonal Walsh Hadamard codes that we used in our interference mitigation approach \cite{key53}. The network consisting of \textit{N} coexisting \textit{WBAN}s that communicate over shared channel where each coordinator is assigned a unique orthogonal spreading code for its \textit{WBAN}. In a time-slot $TS_{i}$ of sensor $r_{i}$ of a \textit{$WBAN_{i}$}, $r_{i}$ multiplies its modulated signal $s_{i}$ by the spreading code $\omega_{i}$. We assume the worst case scenario when $r_{i}$ is interfering with \textit{N-1} sensors in \textit{$TS_{i}$}. The received signal $X_{r}$ at coordinator $C_{i}$ of \textit{$WBAN_{i}$} is given by \textbf{eq.} (\ref{eq1}) below.

\small
\begin{equation}\label{eq1}
X_{r} = \omega_{i}\cdot s_{i} + {\sum_{j=1, j \neq i}^{N-1} \omega_{j}\cdot s_{j} + \mu}
\end{equation}
\normalsize
Inherently the codes generated from the Walsh Hadamard denoted by \textit{WH} matrix $M_{2^n}$ are orthogonal in the zero-phase with \textit{N = n + 1}. $M_{2^n}$ is a special matrix of size $2^N$ $\times$ $2^N$.

\small
\begin{equation}\label{eq:2}
M_{1} =
 \begin{pmatrix}
1
 \end{pmatrix}
 , 
 M_{2} = 
 \begin{pmatrix}
  1 & 1 \\
  1 & -1 
 \end{pmatrix}
\end{equation}
\normalsize
are given, one can generate a generic \textit{WH} matrix $M_{2^n}$, $n > 1$, as follows.

\small
\begin{equation}\label{eq:3}
 M_{2^n} = 
 \begin{pmatrix}
  M_{2^n-1} &  M_{2^n-1}\\
  M_{2^n-1} &  M_{2^n-1} 
 \end{pmatrix}
 = M_{2} \otimes M_{2^n-1}
\end{equation}
\normalsize
Where $\otimes$ denotes the Kronecker product. The rows in each matrix generated in \textbf{eq.} (\ref{eq:3}) are orthogonal to each other. However, the orthogonality property of \textit{WH} codes is lost if the codes are phase shifted. So, a special set of codes extracted from the \textit{WH} matrix $M_{2^k}$ is required to keep orthogonality property with any phase shift ($\phi = 0, 1, 2, \dots 2^{k}-1$). Thus, one can extract \textit{N = n + 1} orthogonal codes from $M_{2^k}$ matrix that have zero cross correlation for all $\phi = 0, 1, 2, \dots 2^{k}-1$. This set of \textit{N} cyclic orthogonal spreading codes is called Orthogonal Walsh Hadamard Codes and denoted by (\textit{COWHC}). If the \textit{COWHC} set is used for spreading in the network, then, $d_{i}$ is the decoded signal of high interfering sensor $r_{i}$ at $C_{i}$, where,

\small
\begin{equation}\label{eq:6}
d_{i} = \omega_{i}\cdot X_{r} = \omega_{i}^2\cdot s_{i} + \sum_{j = 1, j \neq i}^{N-1} \omega_{i}\cdot \omega_{j}\cdot s_{j} + \omega_{i}\cdot  \mu
\end{equation}
\normalsize
$\omega_{i}^2$ = 1 and $\omega_{i}\cdot \omega_{j} = 0$ due to their orthogonality. Therefore, the decoded signal is $d_{i}$ = $s_{i}$ + $\omega_{i}\cdot \mu$.
\section{A Distributed Time Reference Correlation Scheme (\textit{DTRC})}
\begin{figure}
  \centering
        \includegraphics[width=0.325\textwidth]{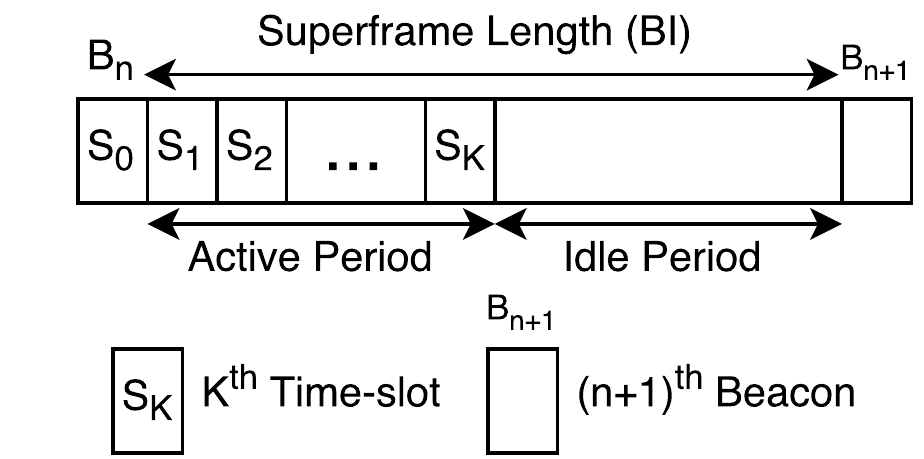}
\caption{Superframe structure proposed for \textit{OCAIM} scheme}
\label{spsfo}
\end{figure}

A \textit{WBAN}'s superframe is delimited by two beacons and composed of equal length active and inactive periods that are dedicated for the sensors and the coordinators, respectively, as shown in \textbf{Figure \ref{spsfo}}.
Due to the absence of inter-\textit{WBAN} coordination and central unit to manage coexistence amongst \textit{WBAN}s, hence, each \textit{WBAN}'s transmission may face collisions with other \textit{WBAN}s' transmissions in the same time-slots. In this work, we do not aim to interleave the superframes or add extra time intervals to avoid collisions. Instead, we present a distributed approach, namely, \textit{DTRC} to avoid such collisions and minimize the time delay of each sensor's transmission. \textit{DTRC} allows each \textit{WBAN} to relate the start time of other superframes to its local time and hence to predict which sensors within its \textit{WBAN} will be interfering with sensors of other \textit{WBAN}s. Thus, all coordinators generate virtual time-based patterns involving the schedule of the transmission and reception of frames. More precisely, each coordinator according to its local clock calculates the \textit{timeshift} from the \textit{actual start transmission time} of a frame. Basically, the \textit{timeshift} comprises, 1) non-deterministic parameters such as the \textit{synchronization error tolerance, the timing uncertainty and the clock drift} and, 2) the \textit{difference} between the \textit{non-deterministic parameters} and the \textit{virtual start transmission time} of a frame \cite{key26}. We define the following parameters that we used in our proposed \textit{DTRC} scheme:
\begin{itemize} 
\item \textit{PHY} Timestamp (\textit{PTP}), encodes the time when the last bit of the frame has transmitted to the air
\item \textit{MAC} Timestamp (\textit{MTP}), encodes the time when the last bit of a frame has been transmitted at the \textit{MAC}
\item \textit{PHY} Receiving Time (\textit{PRT}), a time elapsed from the first to the last bit of a frame at the \textit{PHY}
\item \textit{MAC} Receiving Time (\textit{MRT}), a time elapsed from receiving the first bit to the last bit of a frame at the \textit{MAC}
\item Propagation Delay (\textit{L}), a time elapsed by the bit to travel from the transmitter to the receiver through the air
\item \textit{PHY} Processing Time (\textit{PPT}), a time elapsed from receiving the last bit of a frame at \textit{PHY} until the delivery of the first bit to the \textit{MAC}
\item Frame Reception time (\textit{FRT}), encodes the time when the last bit of a frame has been received at the \textit{MAC}
\end{itemize}
Whenever a coordinator has a frame to transmit, the \textit{MAC} service (resp. the \textit{PHY} service) adds a \textit{MAC}-level timestamp denoted by \textit{MTP} (resp. \textit{PHY}-level timestamp denoted by \textit{PTP}) that encodes the time when the last bit of the frame is transmitted to the \textit{PHY} layer (resp. to the air). Such addition with other \textit{PHY}- and \textit{MAC}-level parameters enable the receiving coordinator to calculate the \textit{timeshift}. Furthermore, when the coordinator receives a frame at the \textit{MAC}, it timestamps the reception of the last bit of that frame through \textit{FRT} according to its local clock. Thus, as the frame bits pass through the \textit{PHY} and \textit{MAC} layers, the receiving services at each layer calculates the following parameters: 1) the time spent by the \textit{MAC} service to receive the frame (\textit{MRT}), 2) the time spent by the \textit{PHY} service to process the frame (\textit{PPT}), 3) the time spent by the \textit{PHY} service to receive the frame (\textit{PRT}) and, 4) the time spent by the first bit of the frame to be received at the \textit{PHY} from the air. 

\begin{figure}
  \centering
        \includegraphics[width=0.325\textwidth]{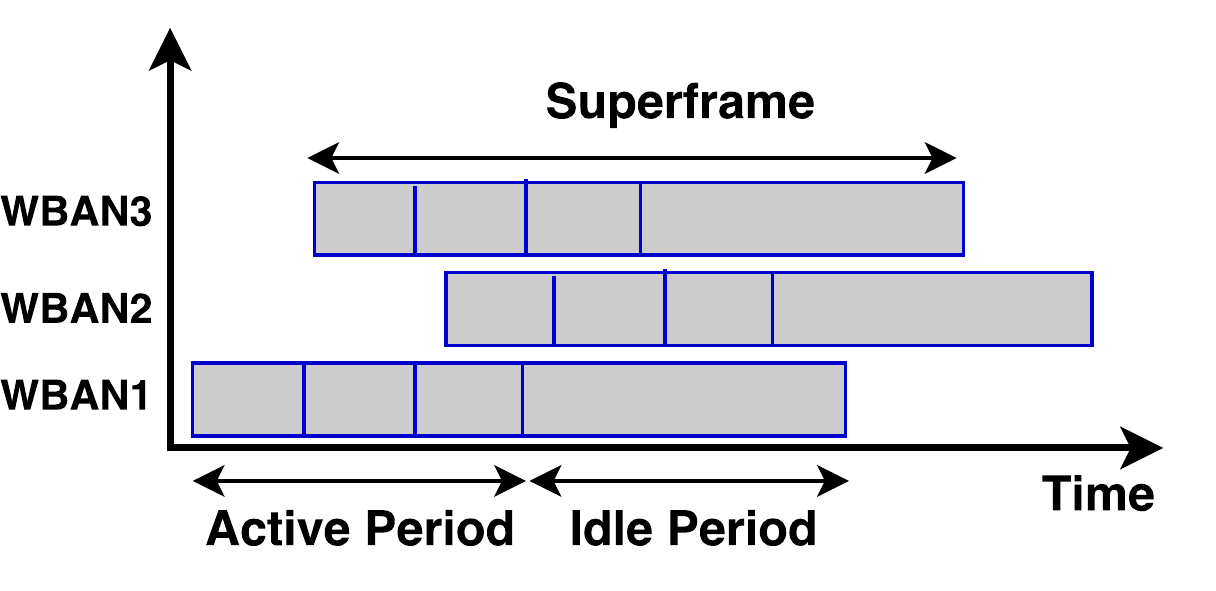}
\caption{Overlapping superframes scheme}
\label{dvsp}
\end{figure}

Subsequently, each coordinator relates the calculated parameters and timestamps as well as the frame reception times to compute the \textit{timeshift} as shown in \textbf{Algorithm \ref{is-iiii}}. Afterwards, it generates a pattern which consists of the different computed \textit{timeshifts} of the different superframes. Based on a \textit{timeshift} of a particular superframe, the coordinator aligns the start transmission time of its superframe to the superframe of that \textit{timeshift} to predict which time-slots within its superframe are interfering with the time-slots of that superframe. To summarize, \textit{DTRC} provides each coordinator with two fundamental functionalities, 1) it determines which superframes may overlap, and more precisely, 2) which time-slots within those superframes may collide with each other as shown in \textbf{Figure \ref{dvsp}}. The pseudocode of \textit{DTRC} is described in \textbf{Algorithm \ref{is-iiii}}.
\begin{algorithm}
\footnotesize
\SetKwData{Left}{left}\SetKwData{This}{this}\SetKwData{Up}{up}
\SetKwFunction{Union}{Union}\SetKwFunction{FindCompress}{FindCompress}
\SetKwInOut{Input}{input}\SetKwInOut{Output}{output}
        
        \Input{N \textit{WBAN}s, K Sensors/\textit{WBAN}, K Slots/Superframe}
       
         \For{i $\leftarrow$ 1  $\KwTo$  N}
          {%
          
              \For{l $\leftarrow$ 1 $\KwTo$ $N-1$ $\&$ i $\neq$ l}
              {%

          $C_{i}$ receives $B^{l}$ at $FRT_{i, l}$ $\&$ $C_{i}$ computes:
          
            $Diff_l = PTP_{l} - MTP_{l} = PPT_{l} + PRT_{l}$   
            
            $timehift_{i,l} = FRT_{i} - [MRT_{i} + PPT_{i} + PRT_{i} + L + Diff_l]$
                   
                    $IntrfrnSlots_{i, l} = timehift_{i,l}/TS$
                    
                    $ID$ =  $\lceil\: \mid IntrfrnSlots_{i, l} \mid\: \rceil$
                  
                  switch ($timehift_{i, l}$)
                  
                  {%
                  
                     case '$timeshift_{i, l}$ $<$ $0$ $\&$ ($\mid timeshift_{i, l}\mid$ $<$ $TS$)':  
    
                      \hspace{0.5cm} ($\forall$ $z$ $\geq$ $ID$ $\&$ $\forall$ $t$ $\geq$ $ID$), $T_{i,z}$ $\bowtie$ $T_{l,t}$
    
                     case '$timeshift_{i, l}$ $<$ $0$ $\&$ ($TS$ $<$ $\mid$ $timeshift_{i, l}$ $\mid$ $<$ $BI/2$)': 
    
                     \hspace{0.5cm} ($\forall$ $z$ $>$ $(K-ID)$ $\&$ $\forall$ $t$ $\leq$ $ID$), $T_{i, z}$ $\bowtie$ $T_{l,t}$
    
                     case '$timeshift_{i, l}$ = $0$':  
        
                  \hspace{0.5cm} Complete interference of $C_{l}$ $\&$ $C_{i}$ active periods 
    
                 }
               
               }

            }

\caption{Proposed \textit{DTRC} Scheme}
\label{is-iiii}
\end{algorithm}
\DecMargin{1em}

\section{Interference Mitigation Using Distributed Orthogonal Codes (\textit{OCAIM})}
\begin{figure}
  \centering
        \includegraphics[width=0.325\textwidth]{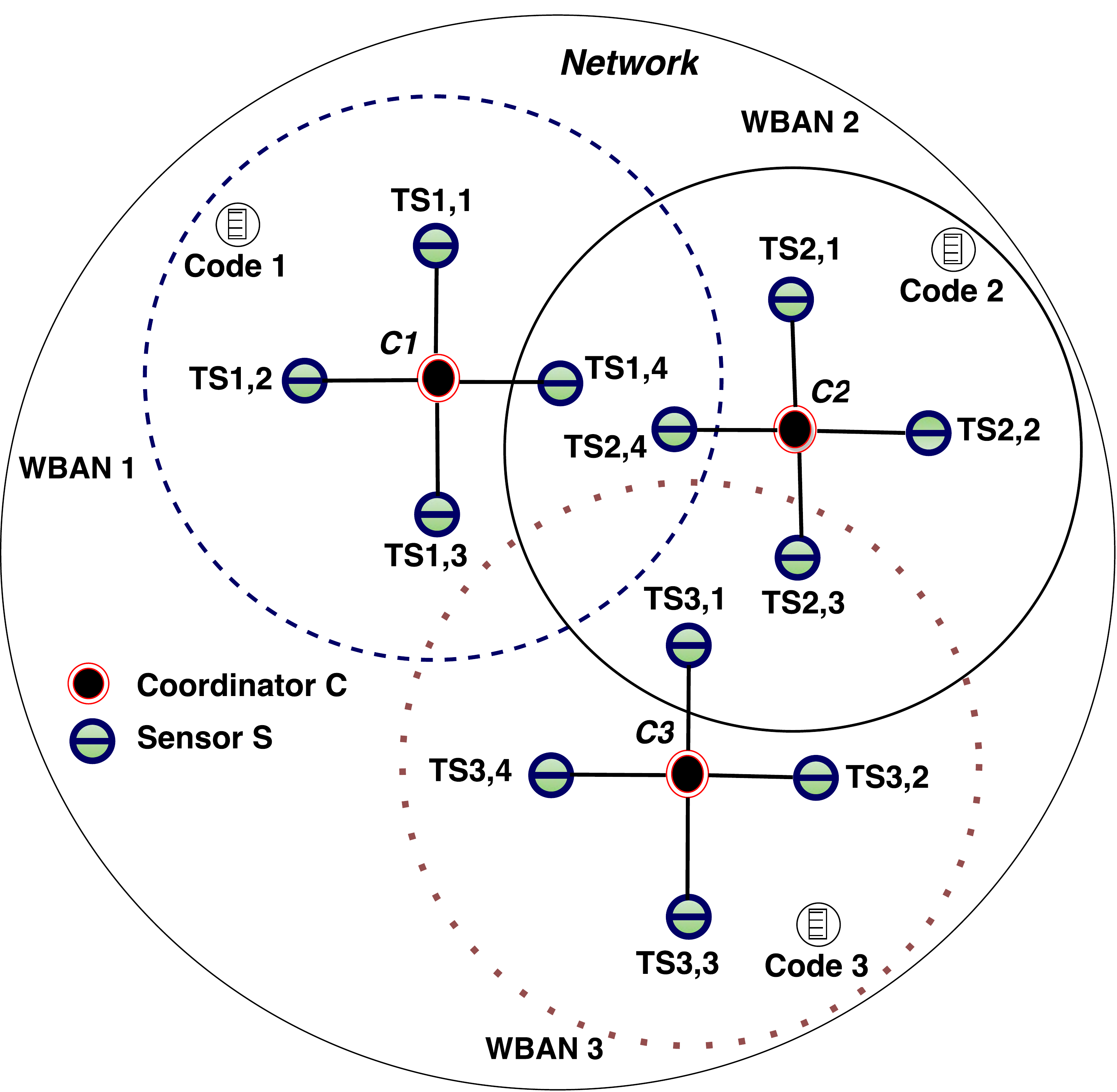}
\caption{A network of three coexisting \textit{WBAN}s}
\label{cdmawban}
\end{figure}

As aforementioned, when two or more sensors of different coexisting \textit{WBAN}s in the same area simultaneously access the shared channel as shown in \textbf{Figure \ref{cdmawban}}, a co-channel interference may arise. Hence, the superframes of different \textit{WBAN}s may overlap as shown in \textbf{Figure \ref{dvsp}}. In our proposed \textit{OCAIM} scheme, each \textit{WBAN} is allocated a unique cyclic orthogonal code from the set \textit{COWHC}. However, based on the interference that \textit{a particular sensor suffers in one or more time-slots it has been assigned, the coordinator commands that sensor to use the code in that time-slots for spreading its signal}. Doing so, each sensor multiplies its signal by a spreading code to increase the bandwidth of that signal and then to become anti-interference. Furthermore, each coordinator updates its code assignment pattern with every new superframe. Additionally, it is important to note that spread spectrum techniques use the same transmit power levels because they transmit at a much lower spectral power density than that of the narrow band transmitters \cite{key26}.

We denote $k^{th}$ \textit{Sensor Interference List} of sensor $S_{i,k}$ of \textit{$WBAN_{i}$} by \textit{$SIL_{i, k}$} that comprises all sensors of other \textit{WBAN}s which impose interference on $S_{i,k}$. Hence, $C_{i}$ adds all sensors $S_{l,m}$ to \textit{$SIL_{i, k}$} that, 1) interfere with $S_{i,k}$ in its assigned time-slot $T_{i,k}$ denoted by $S_{l,m} \bowtie S_{i,k}$ (time-slot level interference is determined by \textit{DTRC}) and, 2) whose binary bitwise OR with that of $S_{i,k}$ equals to 1 denoted by $F_{i,k}$ $\otimes F_{l,m} = 1$, where $F_{i,k}$ and $F_{l,m}$ are indicator functions respectively defined as follows.

\small
\[ F_{i,k} =
  \begin{cases}
    1  & \quad if \; S_{i,k} \in IN_{i, l}\\
    0  & \quad if \; S_{i,k}  \notin IN_{i, l} \\
  \end{cases}
\]
, 
 \[ F_{l,m} =
  \begin{cases}
    1  & \quad if \; S_{l,m} \in IN_{i, l}\\
    0  & \quad if \; S_{l,m} \notin IN_{i, l}\\
  \end{cases}
\]
\normalsize
I.e., \textit{$WBAN_{l}$} is an interferer to \textit{$WBAN_{i}$} and $IN_{i,l} = IS_{i} \cap IS_{l}$. Then, we define $SIL_{i}$ as in \textbf{eq.} (\ref{silik}) below.

\small
\begin{equation}\label{silik}
SIL_{i,k} = \{S_{l,m}|T_{l,m}\bowtie T_{i,k} \: \& \: F_{i,k} \otimes F_{l,m} = 1\}
\end{equation}
\normalsize
Therefore, $C_{i}$ assigns a code to $S_{i, k}$ within its \textit{WBAN} and each sensor belongs to \textit{$SIL_{i,k}$} is also assigned a code within its \textit{WBAN} to avoid the interference. In other words, all interfering sensors of the same \textit{WBAN} use the same code, each in its assigned time-slot since \textit{TDMA} is used within each \textit{WBAN}.

We illustrate our approach through an example of three coexisting \textit{TDMA}-based \textit{WBAN}s scenario as shown in \textbf{Figure \ref{cdmawban}}. However, we denote $j^{th}$ sensor of $\textit{WBAN}_{i}$ is transmitting to its coordinator $C_{i}$ by $S_{i,j}$. Assuming sensors of same index are simultaneously transmitting. The \textit{interference lists} are \textit{$I_{1}=\{S_{2,4}\}$, $I_{2}=\{S_{1,4}, S_{3, 1}\}$, $I_{3}=\{S_{2,3}\}$} and the \textit{interference sets} are \textit{$IS_{1}=\{S_{1,4}, S_{2,4}\}$, $IS_{2}=\{S_{2,3}, S_{2,4}, S_{1,4}, S_{3, 1}\}$, $IS_{3}=\{ S_{3, 1}, S_{2,3}\}$}. Thus, for $\textit{WBAN}_{2}$, the \textit{sensor interference sets} are \textit{$SIL_{2,1}=\{S_{3, 1}\}$, $SIL_{2,2}=\Phi$, $SIL_{2,3}=\{S_{3, 3}\}$} and \textit{$SIL_{2,4}=\{S_{1, 4}\}$}. Then, $C_{2}$ assigns \textit{$Code_2$} to $S_{2,1}$, $S_{2,3}$ and $S_{2,4}$ each in its time-slot, whereas $C_{1}$ assigns \textit{$Code_1$} to $S_{1,4}$ and $C_{3}$ assigns \textit{$Code_3$} to $S_{3,1}$ and $S_{3,3}$. \textbf{Algorithm \ref{is-i}} presents the proposed \textit{OCAIM} scheme.
\begin{algorithm}
\footnotesize
\SetKwData{Left}{left}\SetKwData{This}{this}\SetKwData{Up}{up}
\SetKwFunction{Union}{Union}\SetKwFunction{FindCompress}{FindCompress}
\SetKwInOut{Input}{input}\SetKwInOut{Output}{output}

\Input{N \textit{WBAN}s, K Sensors/\textit{WBAN}}

Phase 1: \textit{TDMA} Orthogonal Transmissions

    \For{i $\leftarrow$ 1  $\KwTo$  N}
       { %
       
            $C_{i}$ broadcasts Beacon $B^{i}$
          
         \For{$k\leftarrow 1$  $\KwTo$  $K$}
          {%
          
            $S_{i,k}$ is transmitting in time-slot $T_{i,k}$ to $C_{i}$
            
            $C_{l}$ $\forall$ $l\neq i$ calculates $\delta_{i,l,k}$
            
          }
          
          $C_{i}$ finds $\rho_{i}^{min}$ = $min \{\delta_{i,k}\}_{\forall k = 1 \dots K}$
          
        } 

Phase 2: Interference Lists (I) and Sets (IS) Formation
      
      \For{$i\leftarrow 1$ $\KwTo$  $N$}
       { %
          
          \For{$l\leftarrow 1$  $\KwTo$  $N$, l $\neq$ i}
            { %
            
             \For{$m\leftarrow 1$  $\KwTo$  $K$}
               {%
               
                  \If{$\delta_{i,l,m}$ > $\rho_{i}^{min}$ - $\theta$}
                   {%
                     
                      Add $S_{l,m}$ to set $I_{i}$
                
                   }
               
                }
                
            }

          $C_{i}$ broadcasts $I_{i}$ $\&$ sets $IS_{i}$ = $I_{i}$ $\cup$ \{$S_{i,k}$ | $S_{i,k}$ $\in$ $I_{l}$, $\forall$ l $\neq$ i\}
          
        }
 
 Phase 3: Distributed Time Reference Correlation Formation (\textit{DTRC})
      
  \For{$i\leftarrow 1$ $\KwTo$ $N$}
  {%
   
   $C_{i}$ executes \textbf{Algorithm \ref{is-iiii}}
  
  }
          
Phase 4: Sensor Interference List (SIL) Formation

  \For{$i\leftarrow 1$ $\KwTo$ $N$}
  {%
              
          \For{$l \leftarrow 1$ $\KwTo$ $N$, $i \neq l$ }
            {%
                  
              $IN_{i,l}$ = \{$IS_{i}$ $\cap$ $IS_{l}$\}
                  
              \For{k $\leftarrow$ 1 $\KwTo$ K}
               {%
                  
                  $SIL_{i,k}$ = \{$(S_{l,m}$ $\mid$ $S_{l,m}$ $\bowtie$ $S_{i,k}$) $\&$ ($F_k$ $\otimes$ $F_m$ = 1)\}
        
               }
        
           }
      
  }

Phase 5: Orthogonal Codes Assignments

  \For{$i\leftarrow 1$ $\KwTo$ $N$}
  {%
  
       \For{$k\leftarrow 1$ $\KwTo$ $K$}
           {%
           
              \For{$l\leftarrow 1$ $\KwTo$ $N$, $ i \neq l$}
               {%
                 
                 \If{$S_{l, m}$ $\in$ $SIL_{i,k}$}
                   {%
                        
                        $C_{i}$ assigns $Code_{i}$ to $S_{i, k}$
                   
                        $C_{l}$ assigns $Code_{l}$ to $S_{l, m}$ 
                        
                   }
        
               }
        
           }
      
  }
 
$C_{i}$ updates $code-to-timeslot-assignment-pattern_{i}$, $\forall i$
 
\caption{Proposed \textit{OCAIM} Scheme}
\label{is-i}
\end{algorithm}
\DecMargin{1em}

\section{\textit{OCAIM} Transmission Probability: Modeling and Analysis}
In this section, we model and analyze the successful and collision probabilities of the beacons and data frames transmissions to validate our approach. For the simplicity of the analysis, we consider all \textit{WBAN}s in the network have similar superframe and time-slot lengths, respectively, denoted by \textit{BI} and \textit{TS}. Basically, a sensor $S_{i}$ transmits multiple data frames separated by short inter-frame spacing (\textit{SIFS}), where each data frame and beacon require transmission time $T_{fr}$ and $T_{B}$, respectively.
\subsection{Successful Beacon Transmission Probability}
We say a superframe does not interfere when its active period is not commencing at the same time when other \textit{WBAN}s are transmitting. If we assume a coordinator succeeds in beacon transmission with a probability $Pr_{succ}$, then a beacon may be lost with probability ($Pr_{lost}$ = 1 - $Pr_{succ}$). We denote the expected number of data frames transmitted by $S_{i}$ during the active period by $Nfrs_{i}$.
However, a sensor $S_{i}$ may occupy the channel for the time duration denoted by $TD_{i}$ or for the whole time-slot, then, $TD_{i}$ per a superframe is calculated in \textbf{eq.} (\ref{td}).

\small
\begin{equation}\label{td}
TD_{i} = Min(TS_{i},Nfrs_{i}\cdot T_{fr} + (Nfrs_{i} - 1)\cdot SIFS)
\end{equation}
\normalsize
The transmission of a beacon may interfere with the transmissions that take place in the active periods of other \textit{WBAN}s, assuming two \textit{WBAN}s coexist, then, the sum of these periods is the duration of possible beacon interference (collision) calculated in \textbf{eq.} (\ref{tbcoll}).

\small
\begin{equation}\label{tbcoll}
T_{Bcoll} = 2\cdot T_{B} + \sum_{i = 1}^{K}(TD_{i} + T_{B})
\end{equation}
\normalsize
Then, the beacon collision probability is calculated in \textbf{eq.} (\ref{prbcoll}).

\small
\begin{equation}\label{prbcoll}
Pr_{Bcoll} = T_{Bcoll}/BI
\end{equation}
\normalsize
Whilst in the case of N coexisting \textit{WBAN}s are collocated, a coordinator may succeed in beacon transmission that does not interfere with the transmission of $N-1$ \textit{WBAN}s. The probability of successful beacon transmission $Pr_{Bsucc}$ is calculated in \textbf{eq.} (\ref{bsucc}) which implies that there will be an expected number $W_{succ}$ \textit{WBAN}s out of $N-1$ \textit{WBAN}s where their beacons and data frames transmissions are successful. $W_{succ}$ is calculated in \textbf{eq.} (\ref{wsucc}).

\small 
\begin{equation}\label{bsucc}
Pr_{Bsucc} = \prod_{i=1}^{N-1}(1-Pr_{Bcoll}) = (1-Pr_{Bcoll})^{N-1}
\end{equation}
\begin{equation}\label{wsucc}
W_{succ} = (N - 1)\cdot Pr_{Bsucc}
\end{equation}
\normalsize
Doing so, \textbf{eq.} \textbf{(\ref{wsucc})} becomes as follows.

\small
\begin{equation}\label{bsuccmodified}
Pr_{Bsucc} = (1-Pr_{Bcoll})^{{(N-1)}\cdot Pr_{Bsucc}}
\end{equation}
\normalsize
\subsection{Successful Data Transmission Probability}
It is interesting to analyze the successful data transmission probability, i.e., the probability of transmitting a data frame successfully without colliding with transmissions of other N-1 \textit{WBAN}s. However, the duration of successful data transmission of each \textit{WBAN} counted on specific periods of the superframe where no collisions take place. This time duration is calculated as in \textbf{eq.} (\ref{ddt}).

\small
\begin{equation}\label{ddt}
D_{succ} = BI\cdot (1-Pr_{Bcoll})^{W_{succ}}
\end{equation}
\normalsize
Similar to (\ref{tbcoll}), the time duration a data frame may collide with the transmission of another \textit{WBAN} will be calculated in \textbf{eq.} {(\ref{dcoll}).

\small
\begin{equation}\label{dcoll}
D_{coll} = \sum_{i=1}^{K}(TD_{i}+T_{fr})
\end{equation}
\normalsize
To present the probability of successful transmission of $\textit{WBAN}_{1}$ coexisting with another $\textit{WBAN}_{2}$, the transmitted data frames of $\textit{WBAN}_{1}$ do not experience collision with the transmitted data frames of $\textit{WBAN}_{2}$ during a time period of $D_{succ} - 2\cdot D_{coll}$ and during the period of $2\cdot D_{coll}$, half of the frames collide on average. The successful probability of $\textit{WBAN}_{1}$ transmission denoted by $Pr_{wbansucc}^{1}$ coexisting with $\textit{WBAN}_{2}$ is calculated as in \textbf{eq.} (\ref{prwbansucc}).

\small
\begin{equation}\label{prwbansucc}
Pr_{wbansucc}^{1} = \frac{D_{succ} - 2\cdot D_{coll}}{D_{succ}}\cdot 1 + \frac{2\cdot D_{coll}}{D_{succ}}\cdot 1/2
\end{equation}
\begin{equation}\label{prwbansucc1}
= (D_{succ} - D_{coll})/D_{succ}
\end{equation}
\normalsize
Moreover, to derive the successful data transmission probability, it is required to know all the data frames generated (G) and the number of data frames successfully transmitted (H) in a superframe. As we mentioned earlier, whenever a beacon is successfully received, $Nfrs_{i}$ frames are expected to be buffered. But, it may or may not be the case that a sensor $S_{i}$ succeed in transmitting all data frames in its assigned time-slot $TS_{i}$ and so the number of frames will be actually transmitted is bounded by the length of its time-slot \textit{TS}. It is calculated as in \textbf{eq.} (\ref{ntxfrs}).

\small
\begin{equation}\label{ntxfrs}
Ntxfrs_{i} = Min(TS/(T_{fr} + SIFS), Nfrs_{i})
\end{equation}
\normalsize
However, a data frame will be successfully transmitted if the beacon received without any collision with other coexisting transmissions. Now, let us calculate the successful data frame transmission probability for sensor $S_{i}$ as in \textbf{eq.} (\ref{prfrsucc}). 

\small
\begin{equation}\label{prfrsucc}
Pr_{FRsucc}^{i} = \frac{H}{G} = \frac{Pr_{Bsucc}\cdot Ntxfrs_{i}\cdot (Pr_{wbansucc}^{1})^{W_{succ}}}{P_{i}}
\end{equation}
\normalsize
By assuming all the beacons are received successfully, this puts an upper bound on the probability of successful data frame transmission. Doing so, the occupancy time of the channel by sensor $S_{i}$ is calculated as follows in \textbf{eq.} (\ref{tdfr}).

\small
\begin{equation}\label{tdfr}
TD_{i} = P_{i}\cdot T_{fr} + (P_{i} - 1)\cdot SIFS
\end{equation}
\normalsize
Similar to \textbf{(\ref{tbcoll})}, the time duration a data frame may collide with the data frames of a coexisting \textit{WBAN} is given by \textbf{eq.} (\ref{dcoll1}). 

\small
\begin{equation}\label{dcoll1}
D_{coll} = \sum_{i=1}^{K}(TD_{i}+T_{fr})
\end{equation}
\normalsize
Moreover, the probability that data frames of $\textit{WBAN}_{1}$ does not collide with the data frames transmissions of $\textit{WBAN}_{2}$ is calculated in \textbf{eq.} (\ref{prfrsucc1}).

\small
\begin{equation}\label{prfrsucc1}
Pr_{FRsucc}^{1} = (BI - D_{coll})/BI
\end{equation}
\normalsize
Whilst this probability is modified to \textbf{eq.} (\ref{prfrsucc2}) below when $\textit{WBAN}_{1}$ coexist with $N-1$ \textit{WBAN}s, i.e., the data frames transmissions of $\textit{WBAN}_{1}$ do not interfere (collide) with the transmissions of $N-1$ coexisting \textit{WBAN}s.

\small
\begin{equation}\label{prfrsucc2}
Pr_{FRsucc} = (Pr_{FRsucc}^{1})^{N-1}
\end{equation}
\normalsize
\section{Simulation Results}
Simulation experiments are conducted to validate the theoretical results and evaluate the performance of the proposed \textit{OCAIM} scheme. Also, a benchmarking is made with smart spectrum allocation \cite{key16} and orthogonal \textit{TDMA} schemes. We have considered variable number of \textit{WBAN}s moving randomly around each others in a space of $5\times 5 \times 5 m^3$, where, each \textit{WBAN} consists of \textit{K = 10} sensors. Additionally, all sensors use the same transmission power at -10 dBm.
\subsection{Signal to Interference plus Noise Ratio (SINR)}
The average SINR versus time for the proposed \textit{OCAIM} and that for the orthogonal \textit{TDMA} \textit{OS} schemes are compared. As can be clearly seen in \textbf{Figure \ref{sinr}}, \textit{OCAIM} achieves more than two times higher SINR (1.5 dB) than \textit{OS} (0.55 dB) and the channel is more stable due to the code assignment to interfering sensors. Consequently, the energy per bit is increased which better makes the signal anti-interference.
\begin{figure}
  \centering
        \includegraphics[width=0.315\textwidth, height=0.2\textheight]{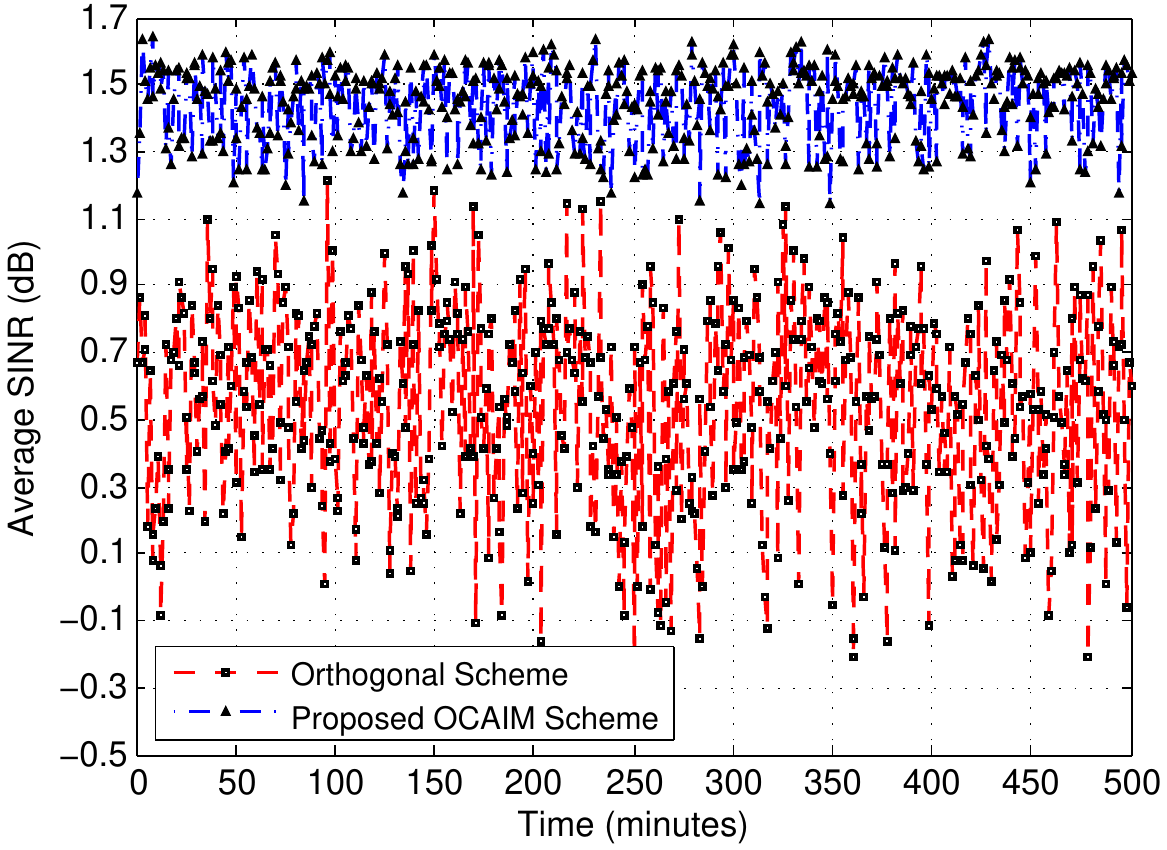}
\caption{Average SINR versus time of \textit{OCAIM} with orthogonal \textit{TDMA} scheme}
\label{sinr}
\end{figure}
\subsection{SINR versus Interference Threshold}
The average SINR versus the interference threshold for \textit{OCAIM} and that for the smart spectrum allocation \textit{SMS} and \textit{OS} schemes are compared. It is observed in \textbf{Figure \ref{interferencesinr}} that SINR of \textit{OCAIM} is higher than that of \textit{SMS} and \textit{OS} for all interference thresholds.
\begin{figure}
  \centering
        \includegraphics[width=0.315\textwidth, height=0.2\textheight]{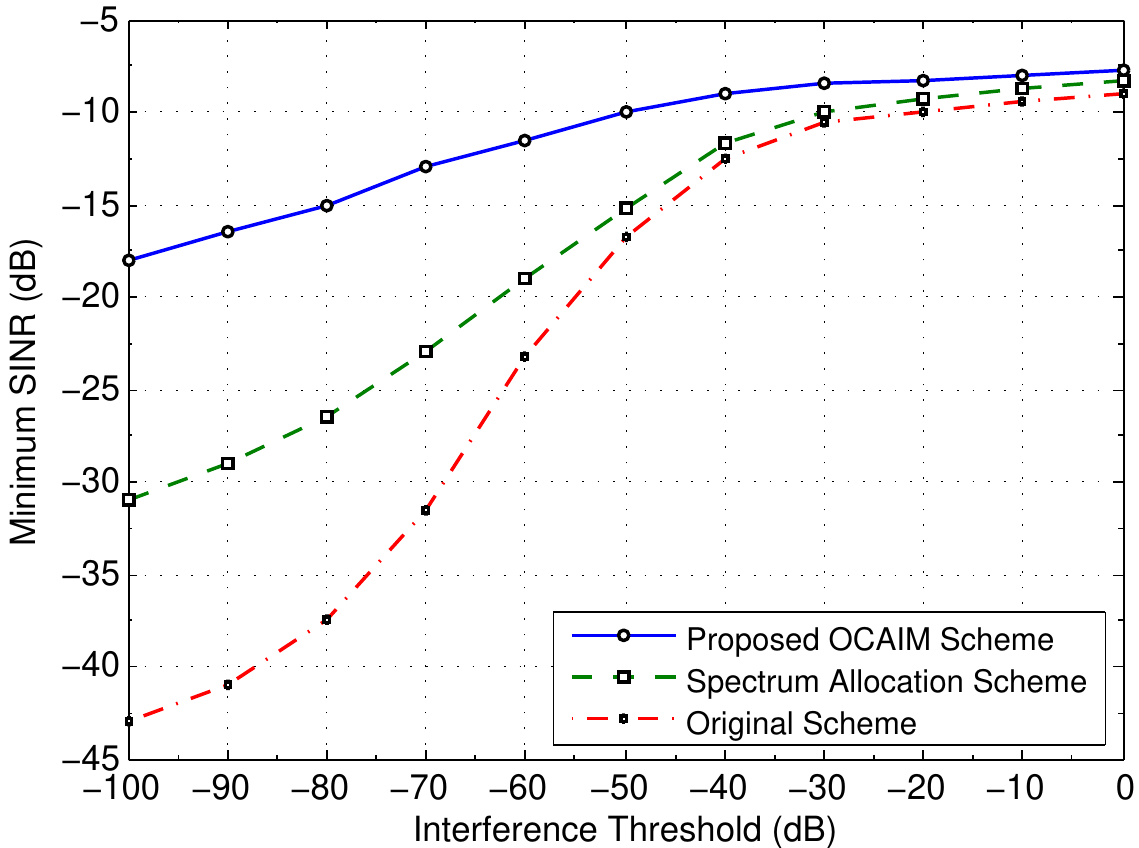}
        \caption{Minimum SINR versus interference threshold of \textit{OCAIM} with \textit{SMS} and \textit{OS} schemes}
\label{interferencesinr}
\end{figure}
However, in \textit{OS}, no coordination is considered (i.e., the probability of superframes overlapping is higher) and neither orthogonal channels nor codes are assigned to the interfering sensors which result in lower values of SINR. On the other side, \textit{OCAIM} considers interference mitigation not only on a sensor-level as in \textit{SMS}, but also on a time-slot level, which explains SINR improvement that \textit{OCAIM} has compared to \textit{SMS}. Furthermore, in all schemes, a higher SINR is achieved when the interference threshold is increased. Thus, decreasing the interference threshold implies more sensors are added to the interference sets (i.e., more sensors are probably assigned orthogonal codes) which lead to higher SINR values. It is improtant to mention that the work in \cite{key16} assigns channels only based on sensor-level interference. In \textit{OCAIM}, codes are assigned and used by sensors only in some particular time-slots where they experience interference, which explains the improvement in the SINR on other competing schemes.
\subsection{WBAN Power Consumption}
The power consumption versus time for \textit{OCAIM} and that for \textit{SMS} and \textit{OS} are compared. However, it is clear from \textbf{Figure \ref{powerconsumption}} that \textit{OCAIM} has lower power consumption ($0.96 \times 10^{-2} mW$) than \textit{SMS} ($1.3 \times 10^{-2} mW$) and \textit{OS} ($1.6 \times 10^{-2} mW$). In \textit{OS}, the overlapping of active periods results in more collisions, which leads to higher power consumption. Whilst, in \textit{SMS}, the coordinators negotiate to assign channels to interfering sensors that justifies the decrease in power consumption compared to \textit{OS}. However, in \textit{OCAIM}, the coordinators still negotiate to assign codes instead of channels, and so, switching the channel to another consumes more power than code assignments which is confirmed by the simulation results shown in \textbf{Figure \ref{powerconsumption}}, and this justifies the increase in power consumption in \textit{SMS}. In addition, \textit{OCAIM} provides a smaller number of sensors that will be assigned codes, which justifies lower consumption than other schemes.
\begin{figure*}
\begin{minipage}[b]{.3075\textwidth}
\centering
\includegraphics[width=1\textwidth, height=0.2\textheight]{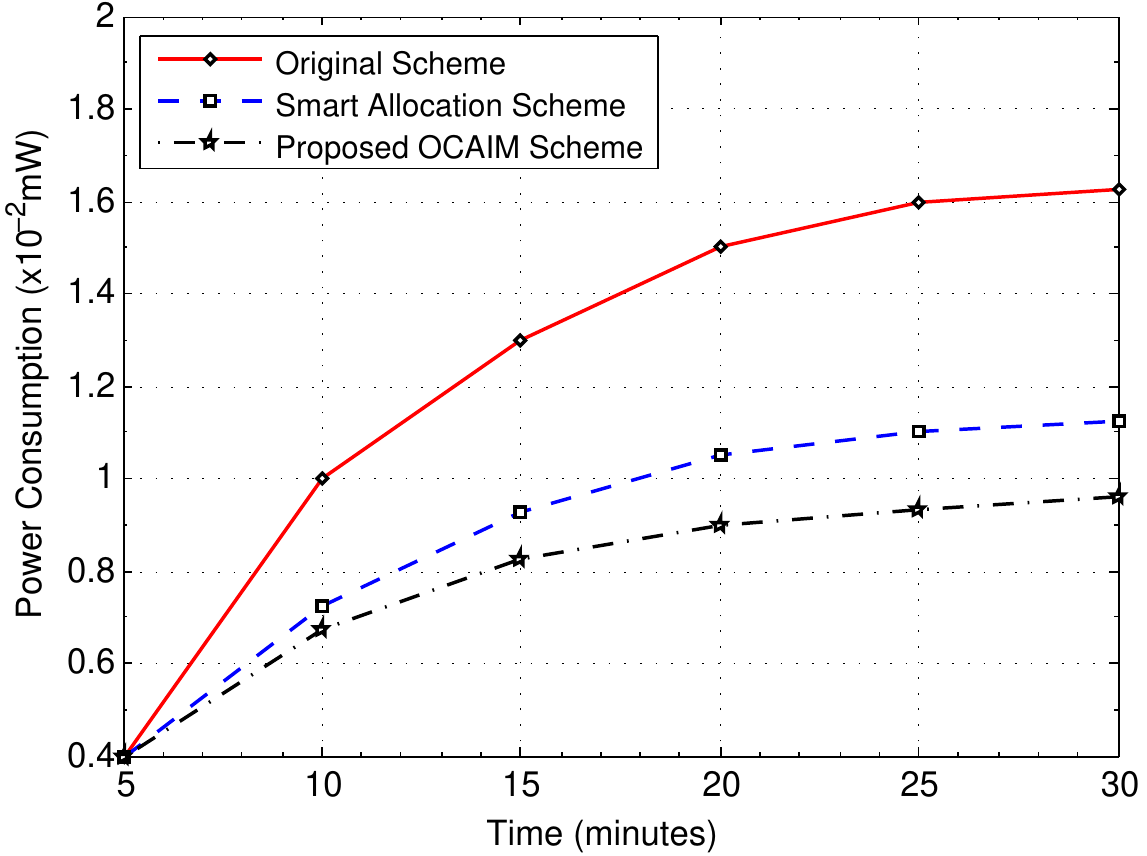}
\caption{\textit{WBAN} power consumption versus time}
\label{powerconsumption}
\end{minipage}\qquad
\begin{minipage}[b]{.3075\textwidth}
\centering
        \includegraphics[width=1\textwidth, height=0.2\textheight]{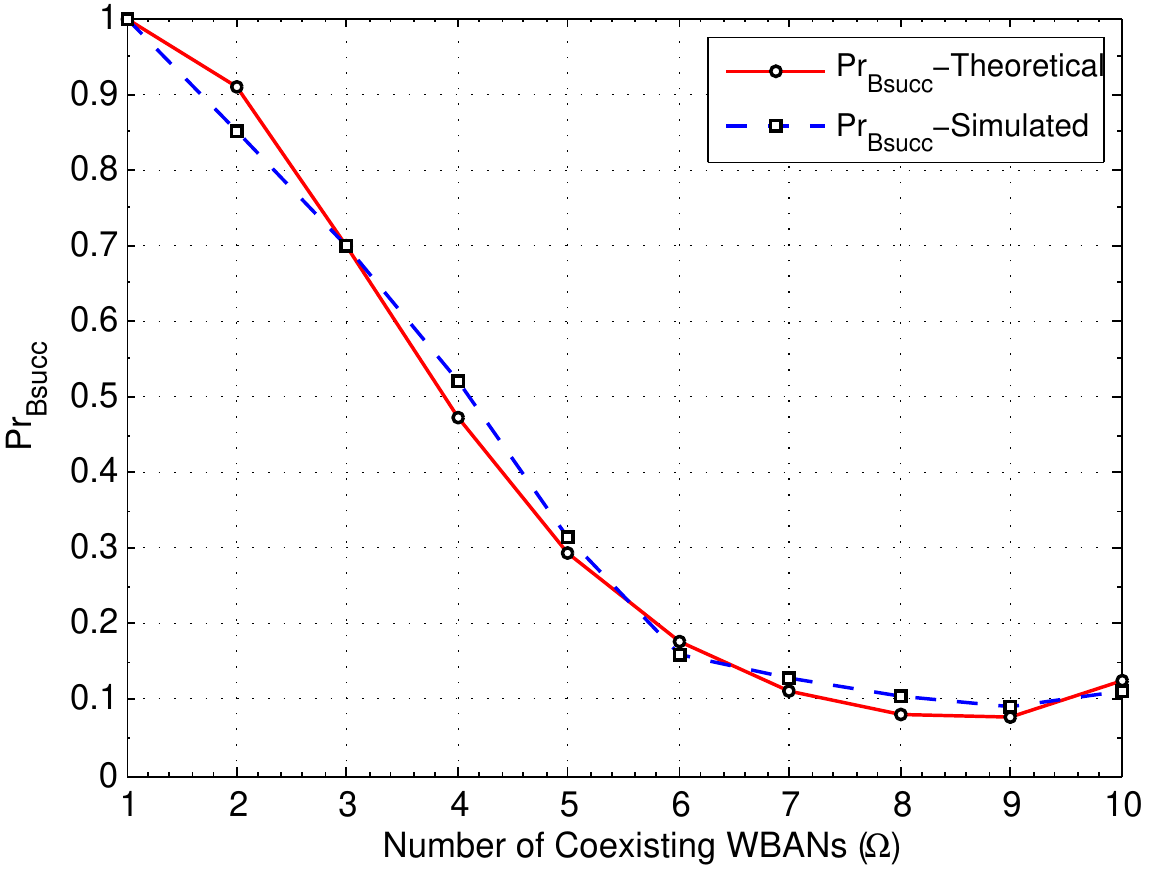}
\caption{Probability of successful beacon transmission versus \textit{WBAN}s count}
\label{prbsucc}
\end{minipage}\qquad
\begin{minipage}[b]{.3075\textwidth}
\centering
        \includegraphics[width=1\textwidth, height=0.2\textheight]{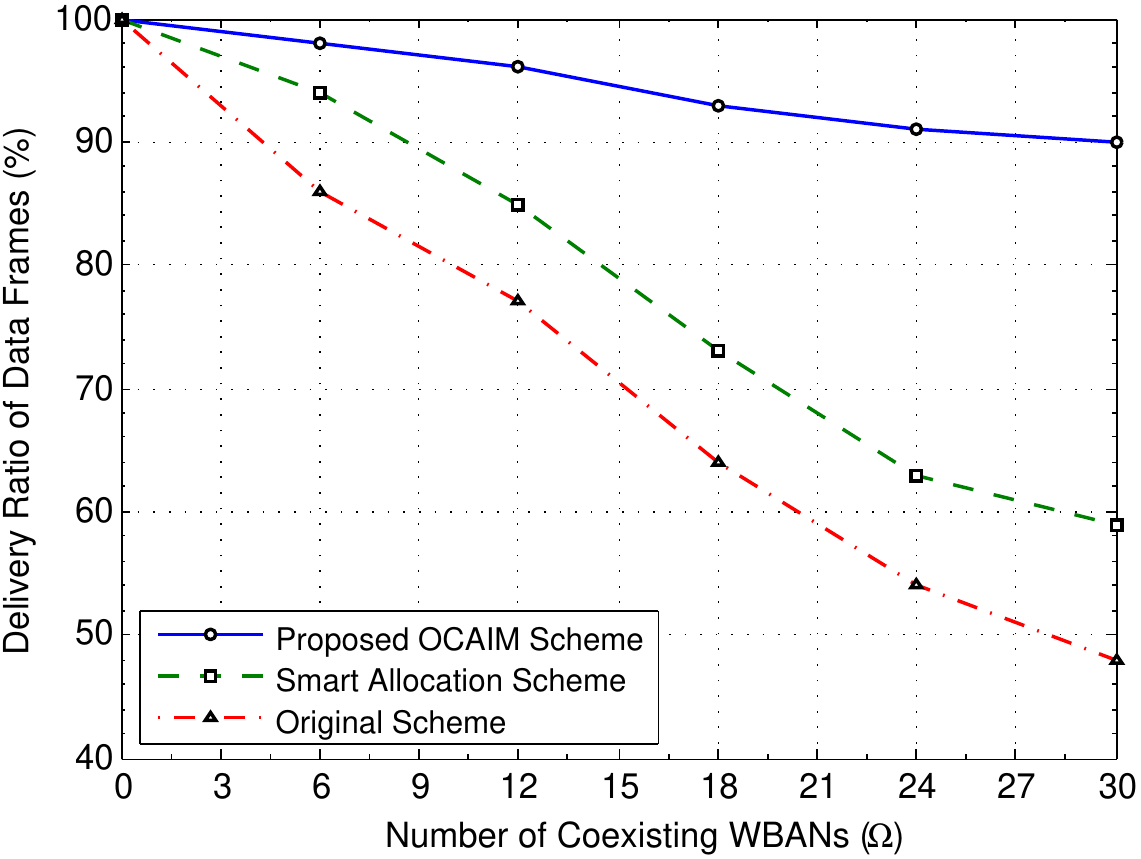}
\caption{Data frames delivery ratio versus \textit{WBAN}s count}
\label{delivery}
\end{minipage}\qquad
\end{figure*}

\subsection{Beacons Transmission Probability}
 \textbf{Figure \ref{prbsucc}} compares the simulated successful beacon transmission probability denoted by $Pr_{Bsucc}^{simulated}$ and the theoretical probability denoted by $Pr_{Bsucc}^{theoretical}$ with varying the \textit{WBAN}s count. As can be clearly seen in this figure, the simulated probability significantly approaches the theoretical one in all cases, which confirms the validity of the theoretical results.
 
\subsection{\textit{WBAN} data frames delivery ratio}
The data frames delivery ratio denoted by \textit{FDR} versus \textit{WBAN}s count $(\Omega)$ for \textit{OCAIM} and that for \textit{SMS} and \textit{OS} are compared. \textbf{Figure \ref{delivery}} shows that \textit{FDR} of \textit{OCAIM} is always higher than that of \textit{SMS} and \textit{OS} for all values of $\Omega$. However, in \textit{OS}, the overlapping of active periods results in more collisions due to the absence of coordination and orthogonal channel/code assignments, which leads to lower values of \textit{FDR}. Whilst, in \textit{SMS} where the number of channels is limited to 16, the coordinators negotiate to assign channels to interfering sensors that justify the increase in \textit{FDR} compared to \textit{OS}. Furthermore, the work in \cite{key16} assigns channels only based on sensor-level interference. However, in \textit{OCAIM}, codes are assigned to sensors only in some particular time-slots where they experience high interference which explains the improvement in \textit{FDR} on other competing schemes. 
\section{Conclusions}
In this paper, a distributed orthogonal code allocation scheme is proposed to avoid co-channel interference amongst coexisting \textit{WBAN}s. To the best of our knowledge, we are the first that consider the interference at the sensor- and time-slot- levels. In our proposed scheme, all the sensors and in their assigned time-slots, where they only impose high interference on other \textit{WBAN}s are allocated orthogonal codes, whilst, other sensors are not required to be assigned codes all the time. Furthermore, the proposed scheme mitigates the interference and increases the power savings at sensor- and \textit{WBAN}-levels, as well as efficiently utilizes the limited resources in \textit{WBAN}s. The performance has been evaluated by extensive experiments and results show the proposed scheme outperforms other competing approaches in terms of interference and power consumption. 


\end{document}